\documentclass[twocolumn,letterpaper,showpacs,floatfix,aps]{revtex4}

\usepackage[pdftex]{graphicx}
\usepackage{amssymb}
\usepackage{esint}
\usepackage{hyperref}

\def \beq {\begin{equation}}
\def \eeq {\end{equation}}

\begin{document}

\title{Doughnut-shaped soap bubbles}

\author{Deison Pr\'eve}
\author{Alberto Saa}\thanks{Corresponding author}
\email{asaa@ime.unicamp.br}
\affiliation{
Departamento de Matem\'atica Aplicada,
IMECC -- UNICAMP, 13083-859 Campinas, SP, Brazil.}


\begin{abstract}
Soap bubbles are thin liquid films
enclosing a fixed volume of air. Since the surface tension is typically assumed to be the only responsible for
conforming the soap bubble shape, the realized bubble surfaces are always minimal area ones. Here, 
we consider   the problem of finding the axisymmetric minimal area surface enclosing a fixed
volume $V$ and with a fixed equatorial perimeter $L$. It is well known that the sphere is the solution for 
$V=L^3/6\pi^2$, and this is indeed the case of a free soap bubble, for instance. 
 Surprisingly, we show that for $V<\alpha L^3/6\pi^2$, with $\alpha\approx 0.21$, such a surface cannot be
  the usual lens-shaped surface formed by the juxtaposition of two spherical caps, but rather a toroidal surface. Practically,
a doughnut-shaped bubble  is known to be ultimately unstable and, hence, it will eventually lose its axisymmetry by breaking apart in smaller 
bubbles. Indisputably, however, the topological transition from spherical to toroidal surfaces is mandatory here for obtaining the
global solution for this
axisymmetric isoperimetric problem. Our result suggests  that deformed bubbles with  $V<\alpha L^3/6\pi^2$ cannot be stable and
should not exist in foams, for instance.  
\end{abstract}

\maketitle

\section{Introduction}
Soap bubbles have been attracting  the attention of physics and mathematicians for more than two centuries \cite{Soap}. A soap bubble is
a thin liquid film enclosing a given volume of air. Surface tension is usually assumed to be
the only responsible for  conforming the bubble surface shape, 
and hence 
 the realized surfaces are always minimal area ones.
It is well known that the sphere is the solution for one of the most celebrated isoperimetric problems:
to find the minimal area surface enclosing a fixed and given volume. Free soap bubbles are known to be spheres.  

\begin{figure}[ht]
\begin{center}
\includegraphics[scale=0.5]{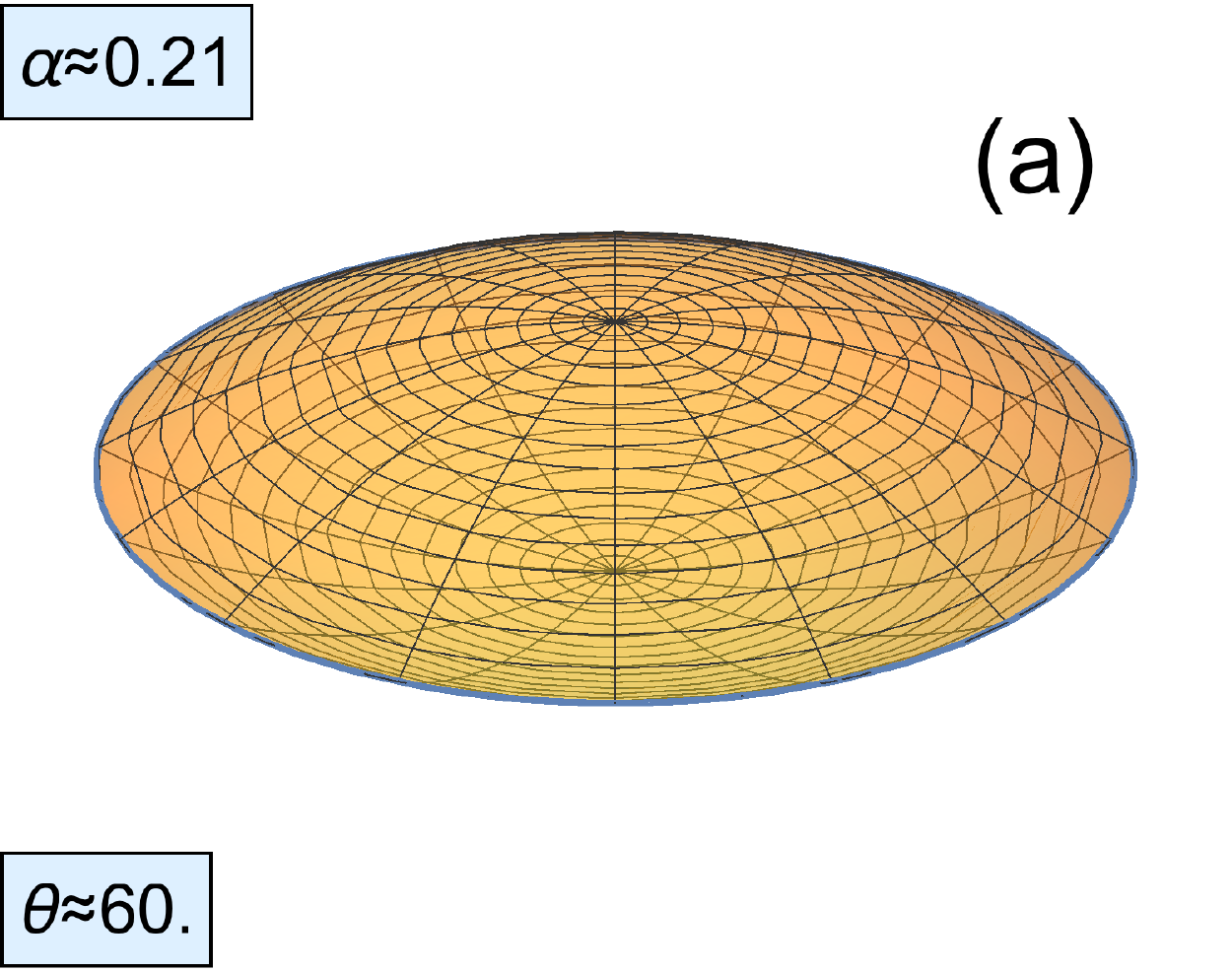}
\vspace{0.5cm}\\
\includegraphics[scale=0.5]{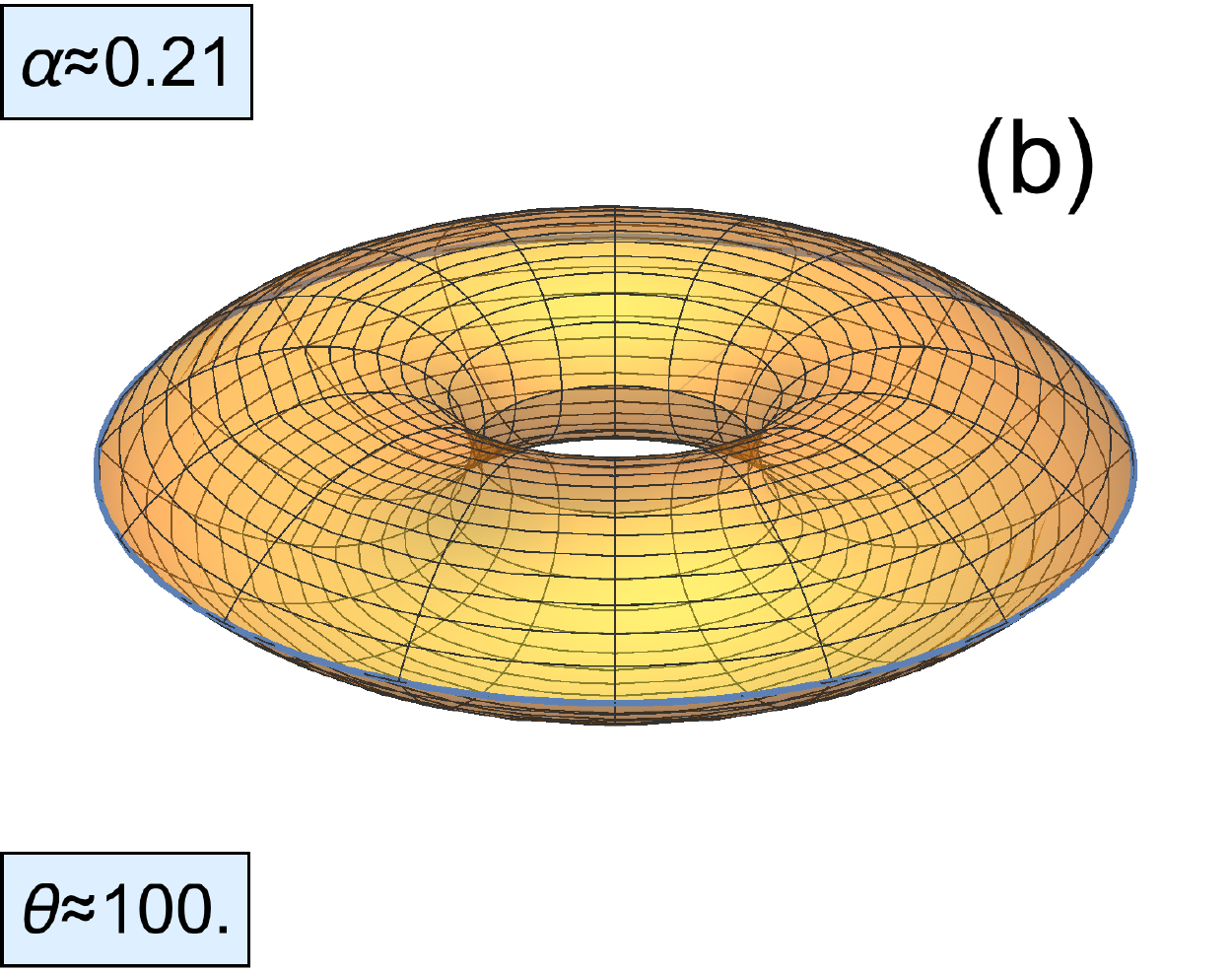}
\end{center}
\caption{Top: lens-shaped surface of minimal area with perimeter $L$ and volume $V=\alpha L^3/6\pi^2$, with $\alpha\approx 0.21$. No stable lens-shaped
surface with $V<\alpha L^3/6\pi^2$ should exist. Bottom: doughnut-shaped surface of minimal area with perimeter $L$ and $V=\alpha L^3/6\pi^2$.
Axisymmetric minimal area surfaces with $V<\alpha L^3/6\pi^2$ are necessarily of this type. No doughnut-shaped minimal area surfaces exist with $V>\alpha L^3/6\pi^2$. In both cases, the angle   $\theta$ is the internal angle of the surface at the equatorial perimeter. 
}
\label{fig0}
\end{figure}

 We consider here the problem
of finding the axisymmetric
minimal area surface with two simultaneous constraints: a fixed enclosed volume $V$ and a fixed equatorial perimeter $L$. Since a sphere
of radius $a$ is the minimal area surface enclosing a volume $V=4\pi a^3/3$, it will be also the solution for our problem
for this volume and  equatorial perimeter $L=2\pi a$. We are mainly interested in the cases with $L=2\pi a$ and $V\le 4\pi a^3/3$, for which the solutions may have the shape of
a ``lens'' formed by the juxtaposition of two spherical caps of height $h < a$, see Fig. \ref{fig0}(a).
The enclosed volume by these lens-shaped surfaces are
\begin{equation}
\label{volu}
V = \frac{\pi h}{3}\left(3a^2+h^2\right),
\end{equation}
whereas their surface area is given by
\begin{equation}
\label{area}
A =  2\pi \left(a^2+h^2\right).
\end{equation}
As we can see, for a fixed equatorial perimeter $L=2\pi a$, one can effectively  have arbitrarily small enclosed volumes $V$
 by choosing arbitrarily small
cap heights $h$ since $0\le V \le L^3/6\pi^2 $. On the other hand, the surface area $A$  will be always bounded from below by a positive value, $A> L^2/2\pi$. It is clear that for small
cap height $h$, the lens-shaped surface cannot be an efficient surface for enclosing a small volume $V$ with a fixed equatorial perimeter $L$.
Surprisingly, in order to obtain the global minimum for this axisymetric  isoperimetric problem, a topological transition is mandatory: from the spherical lens-shaped to toroidal
surfaces. As we will show, lens-shaped surfaces of equatorial perimeter $L$ are global solutions for our minimal area problem only for volumes $V$ such that
\beq
\label{cond}
\alpha \frac{L^3}{6\pi^2} <  V \le  \frac{L^3}{6\pi^2},
\eeq
with $\alpha\approx 0.21$. For $V<\alpha L^3/{6\pi^2}$, the axisymmetric minimal area surface enclosing a volume $V$ will be 
necessarily doughnut-shaped as that one depict in Fig. \ref{fig0}(b), as we
will see by considering all solutions of our
isoperimetric problem in the following section.

\section{The isoperimetric variational problem}

Strictly speaking, the isoperimetric problem, dating from the antiquity, 
concerns finding the plane figure of maximal area  with a given perimeter. In a
broader sense, however, it includes, for instance, the problem of finding the function $f(x,y)$ which
minimize a given functional
\beq
\label{e1}
S[f] = \iint_{\cal D} {\cal L}(x,y,f,f_x,f_y)\; dxdy,
\eeq
but subjected to integral  constraints of the type
\beq
\label{e2}
  \iint_{\cal D} {\cal C}(x,y,f,f_x,f_y) \; dxdy = {\rm constant,}
\eeq
where ${\cal D}$ is a region of the plane $(x,y)$ and the indices $x$ and $y$ denote the respective partial derivatives.
Every function here is assumed to be real and smooth. For our purposes,
let us consider the surface $(x,y,z)$ defined by the function $f(x,y) = \pm z $, which we will assume to be non-negative and such
that $f(x,y)=0$ for $(x,y)\in \partial {\cal D}$.

  The standard
treatment for the isoperimetric problems involves the associated functional defined as
\beq
\label{e3}
S^*[f] = \iint_{\cal D} {\cal L}^* \; dxdy,\quad {\cal L}^*  =
 {\cal L}  + \lambda  {\cal C} ,
\eeq
where $\lambda$  is a constant (the Lagrange multiplier). The function $f(x,y)$ that extremizes (\ref{e1}) 
subjected to the constraint (\ref{e2}) also extremizes the
free functional (\ref{e3}), {\em i.e.}  $f(x,y)$ is a solution of the Euler-Lagrange
equations for the associated functional (\ref{e3}). The constant $\lambda$ is to be determined,
among all the integration constants, from the boundary conditions and the integral constraint
(\ref{e2}). For the  problem of the minimal area axisymmetric surface, one can introduce  
appropriate polar coordinates $(\rho,\theta)$ centred at the surface such that 
  $f=f(\rho)$, 
leading to the following expression for the
area functional
\beq
S[f] = 2\pi \int_{\cal D} \sqrt{1+{f'}^2}\, \rho d\rho,
\eeq
while the fixed volume constraint   will read simply
\beq
\label{vol}
V[f] = 2\pi \int_{\cal D} f \,\rho d\rho = {\rm constant}.
\eeq
Clearly, the equatorial perimeter will be given by the length of $\partial {\cal D}$.
In these coordinates, the Euler-Lagrange equation for the associated functional (\ref{e3}) is
\beq
\label{el}
\frac{1}{\rho}\frac{d}{d\rho}\left(
\frac{\rho f'}{\sqrt{1+{f'}^2} } \right)= \lambda ,
\eeq
which be easily integrated and leads to
\beq
\label{el1}
\frac{\rho f'}{\sqrt{1+{f'}^2} } = \frac{\lambda}{2}\rho^2 + C_1.
\eeq
We have two qualitative distinct cases according to the value of $C_1$.
For $C_1=0$, we have  $f'(0)=0$, which is indeed a regularity condition
for axisymmetric surfaces. However, and more importantly, in this case there is no restriction
for the values of $\rho$ and, consequently, ${\cal D}$ is a circle.   
It is quite simple to verify that the solutions of (\ref{el1}) for this case are the   arcs given by 
\beq
\label{arc}
(f(\rho) + b)^2 + \rho^2 = r_0^2,
\eeq
with $r_0=2/\lambda$ and $r_0>b\ge 0$. These solutions corresponds to the usual spherical cap with basis radius
$a^2 = r_0^2-b^2$ and height $h=r_0 - b$. These caps form the lens-shaped solutions for our isoperimetric problem.

\begin{figure}[ht]
\begin{center}
\includegraphics[scale=0.7]{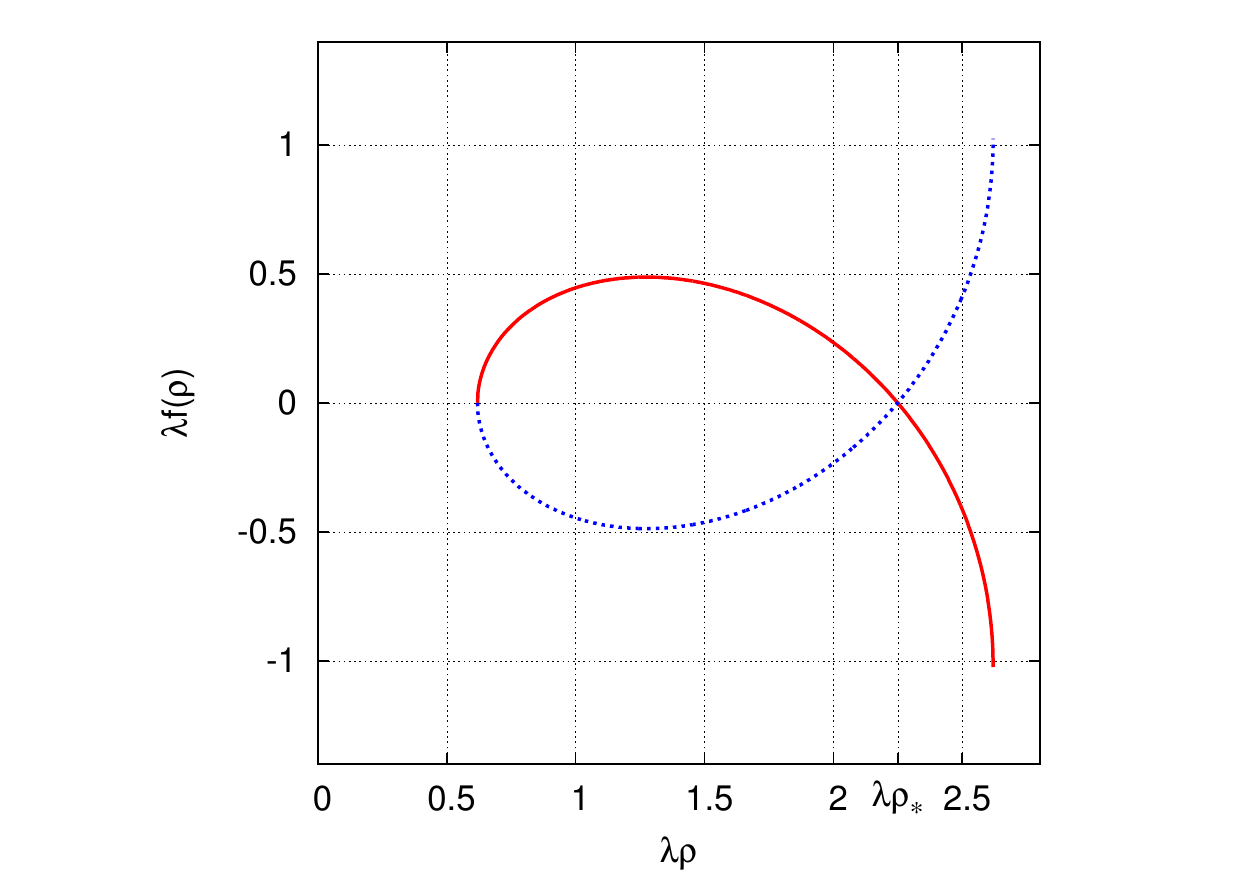}
\end{center}
\caption{Solid red line: solution for (\ref{e15}) with boundary condition $f(\rho_{\rm min})=0$  for $d\approx 1.6235$, which corresponds to  
 $\lambda\rho_{\rm min}\approx 0.6197$, $\lambda\rho_*\approx 2.2491$, and $\lambda\rho_{\rm max}\approx 2.6197$.  The doughnut-shaped solution 
 depicted in \ref{fig0}(b) is obtained by the revolution around the
vertical axis of the closed curve formed by the solution and its reflection on the horizontal axis (dashed blue line) for
$\rho_{\rm min}\le\rho \le \rho_*$. This particular value of $d$ corresponds to the toroidal solution with maximal enclosed volume. The
doughnut-shaped surface is regular everywhere except on the equatorial perimeter $\rho=\rho_*$.}
\label{fig1}
\end{figure}

Nevertheless, we have also the solutions with $C_1\ne 0$. Solving (\ref{el1}) for $f'$ and considering the convenient
signs, we have
\beq
\label{e15}
f'(\rho)   = 
\frac{d - x^2}
{\sqrt{(x^2-x^2_{\rm min})(x^2_{\rm max}-x^2)}},
\eeq
where $x=\lambda \rho$, $d = 2\lambda C_1>0$, 
and 
\beq 
\label{x}
 x_{\rm min}  = \sqrt{1+d} -1, \quad  x_{\rm max} = \sqrt{1+d} +1 .
\eeq 
The first observation here is the most important one: for $C_1\ne 0$,
there will be necessarily restrictions for $\rho$, we have indeed $x_{\rm min} \le \lambda\rho \le x_{\rm max}$. The region
${\cal D}$ is not anymore a circle, but  effectively a ring domain. The integral (\ref{e15}) can be solved analytically by using
elliptic functions, but for our purposes we opt to solve it numerically. All pertinent details are presented in the Appendix. 
The second additive integration constant will
be chosen in order to have $f(\rho_{\rm min})=0$, with $x_{\rm min} = \lambda\rho_{\min}$. Notice that $f'(\rho_{\rm min})$
diverges, assuring in this way that the juxtaposition of the superior and inferior parts of our doughnut-shaped surface will be indeed smooth along
the interior radius. 
An example of solution $f(\rho)$ is depicted in Fig. \ref{fig1}, which corresponds to he doughnut-shaped surface in Fig. \ref{fig0}(b).
 The equatorial perimeter of the surface will be given by
$L=2\pi\rho_*$, where the
radius   $ \rho_*  > \rho_{\rm min}$ is such that $f(\rho_*)=0$.  

The constant $\lambda$ can be effectively absorbed by a global rescaling. For each value of $d>0$, we have a toroidal surface. From (\ref{x}), we see
that small values of $d$ correspond
to the cases with $x_{\rm min}\approx 0$ and $x_{\rm max}\approx 2$. These solutions can enclose arbitrarily small volumes, but their
area is bounded from below by the area of a lens-shaped solution with same equatorial perimeter $L$. 
\begin{figure}[ht]
\begin{center}
\includegraphics[scale=0.7]{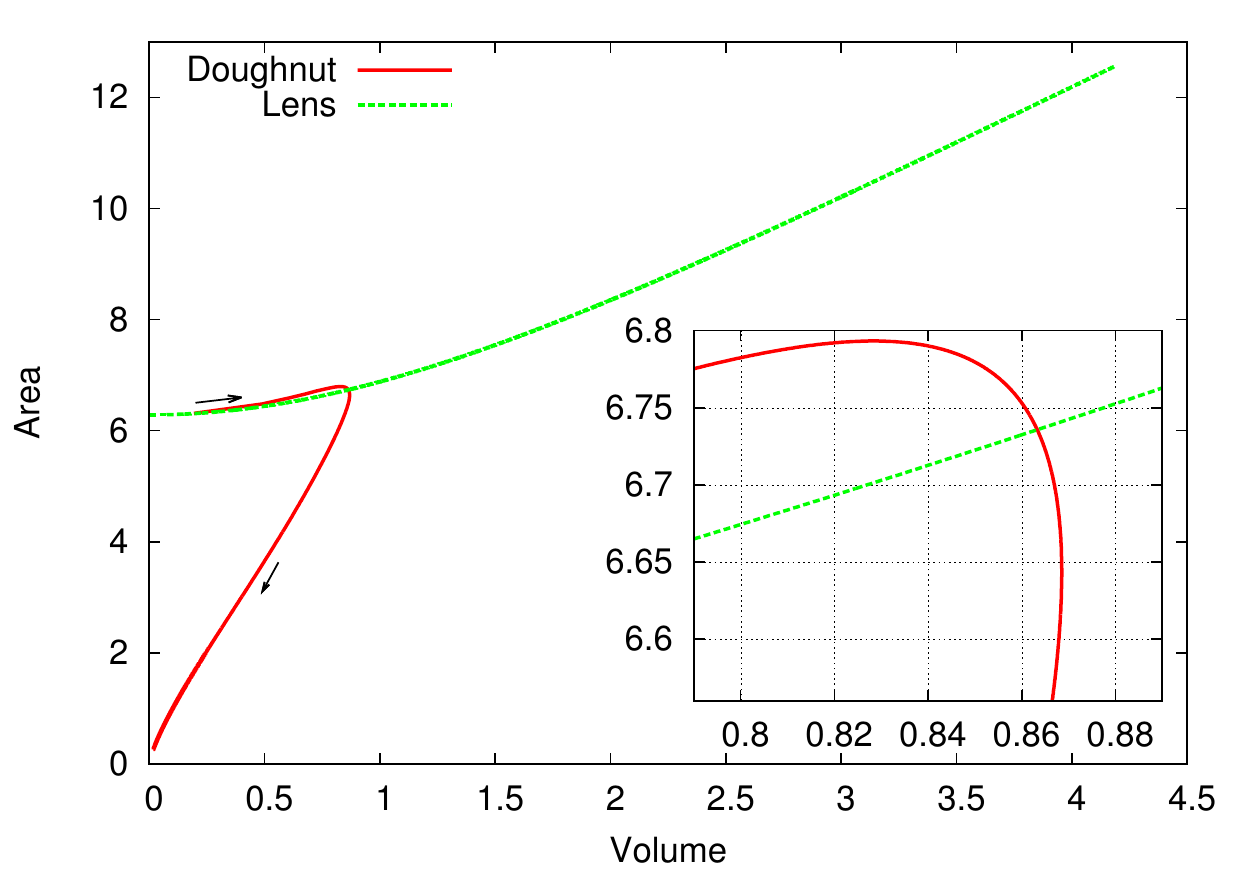}
\end{center}
\caption{Area $\times$ Volume diagram for axisymmetric minimal area surface with fixed equatorial perimeter $L=2\pi$. The solid red line
corresponds to the doughnut-shaped solutions, with the arrows indicating the direction of increasing $d$. The dashed green line corresponds
to the lens-shaped solutions. In the detail, the region corresponding to the topological transition. The maximum volume for the
doughnut-shaped solution is $V\approx 0.869$, corresponding to the case depicted in Fig. \ref{fig1}. An animation illustrating the transition from spherical to toroidal surfaces is available at \cite{anim}. }
\label{fig2}
\end{figure}
On the other hand, the solutions with large $d$, which corresponds to large $x_{\rm min}$ and $x_{\rm max}$, can enclose arbitrarily small volumes
with arbitrarily small surface areas. The situation is depicted in the Area $\times$ Volume diagram of Fig. \ref{fig2}. The solid red line 
corresponds to the doughnut-shaped solutions, while lens-shaped ones correspond to the the dashed green line. The doughnut-shaped solution
with maximal volume corresponds to the case with  $d\approx 1.6235$ (depicted in Fig. \ref{fig0}(b) and Fig. \ref{fig1}). For any other value of $d$, there are
always two minimal area surfaces: one corresponding to  small  $x_{\rm min}/x_{\rm max}$ (small $d$), and the other to  small   
$(x_{\rm max}-x_{\rm min})/x_{\rm max}$ (large $d$). The second one will be the the global minimum of the problem, see Fig. \ref{fig2}. 
We see from the diagram
that the lens-shaped surfaces are effectively the only minimal area solution for our problem provided that the condition (\ref{cond}) holds,
with $\alpha\approx 0.21$, which corresponds namely to the minimal area doughnut-shaped surface of maximum volume. For $V<\alpha L^3/{6\pi^2}$,
we see from the diagram that three minimal area surfaces coexist, but clearly the global minimum corresponds to the case of
doughnut-shaped surfaces with    $d> 1.6235$. An animation illustrating the topological transition from spherical to toroidal surfaces
according to
the value of the ratio $V/L^3$  is available in the Supplementary Material.

\begin{figure}[ht]
\begin{center}
\includegraphics[scale=0.7]{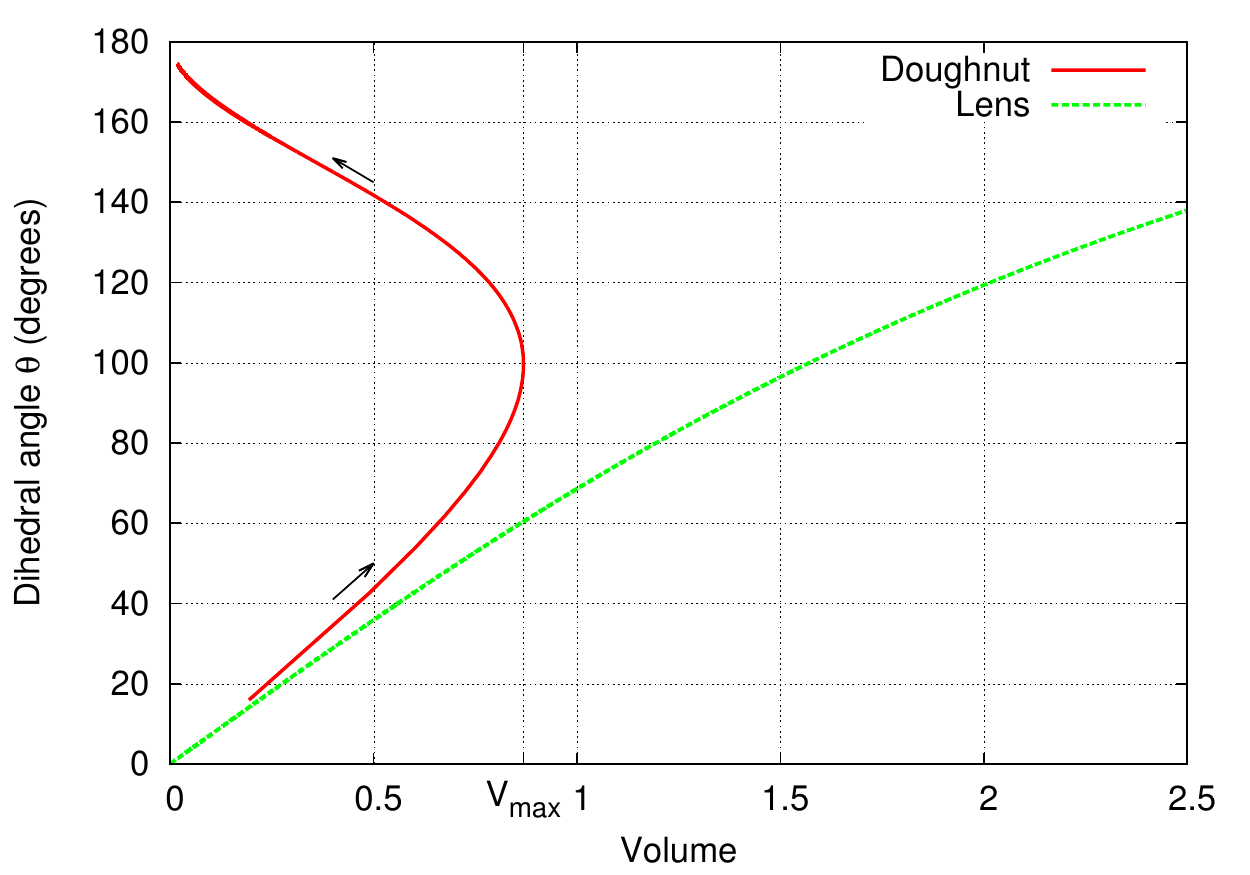}
\end{center}
\caption{ Dihedral angle $\theta$ between the tangent planes
at the equatorial perimeter as a function of the enclosed volume $V$ for axisymmetric minimal area surfaces with equatorial
perimeter $L=2\pi$. The solid red line
corresponds to the doughnut-shaped solutions, with the arrows indicating the direction of increasing $d$, while
the lens-shaped solutions are the dashed green line. Notice that for $d\to \infty$, $V\to 0$ and the dihedral angle
tends to 180 degrees. The doughnut-shaped surface in this limit tends to the usual torus of circular section.}
\label{fig3}
\end{figure}

It is interesting to relate the topological transition of the minimal area surfaces to the dihedral angle $\theta$ between the tangent planes
at the equatorial perimeter. Fig. \ref{fig3} depicts the dependence of $\theta$ on the volume $V$ for a fixed equatorial perimeter
$L=2\pi$.
For the lens-shaped surfaces, the minimal volume $V=\alpha L^3/6\pi^2$ case corresponds to $\theta$ close
to $60$ degrees. This is the case depicted in Fig. \ref{fig0}(a).  The doughnut-shapes surface with the same volume has a larger dihedral angle, close to 
$100$ degrees  (Fig. \ref{fig0}(b) and Fig. \ref{fig1}). Notice that
\beq
\alpha = 8 -\frac{9}{2}\sqrt{3} = 0.20577\dots
\eeq
corresponds to the volume of
  a  lens-shaped surface with dihedral angle $\theta = 60$ degrees at the equatorial external perimeter. 

Notice that all solutions to our isoperimetrical problem are indeed constant mean curvatures. This can be checked by recalling the first
and second, respectively, fundamental forms for our surface of revolution generated by $f(\rho)$: $E=\rho^2$, $F=0$, $G=1+ {f'}^2 $; and
$L=-\rho f'/\sqrt{1+{f'}^2}$, $M=0$,  $N=-f''/\sqrt{1+{f'}^2}$; and that
\beq
-\frac{1}{\rho}\frac{d}{d\rho}\left(
\frac{\rho f'}{\sqrt{1+{f'}^2} } \right) = \frac{L}{E}+\frac{N}{G} =2H,
\eeq
where $H$ stands for the mean curvature of our surface. Thus, the Euler-Lagrange equation (\ref{el}) is equivalent to the constraint
of constant $H$. Since $H$ is constant, we can easily evaluate it by taking the point $\bar\rho$ such that $f'(\bar\rho)=0$, where $L=0$ 
and $N=-f''(\bar\rho)$, leading simply to 
\beq
H = -\frac{1}{2}f''(\bar\rho),
\eeq
which is positive for our toroidal surfaces. 
The mean curvature can be expressed also as 
 $2H= R_1^{-1}+R_2^{-1}$, where $R_1$ and $R_2$ are    the radii corresponding to the principal curvatures. The  $d\to \infty$ ($V\to 0$) limit of figure (\ref{fig3}), for which  
 the external dihedral angle
tends to 180 degrees, is an usual torus of circular section for which $R_1\ll R_2$, assuring in this way that $2H\approx R_1^{-1} =$ constant.

\section{Final remarks}

We have shown that the isoperimetric problem of finding an asixymmetric minimal area surface enclosing a fixed volume $V$ and with
a fixed equatorial perimeter $L$ exhibits a rather unexpected topological transition in their solutions accordingly to the ratio
$V/L^3$. The typical lens-shaped surfaces formed by
two spherical caps are not the global minimum for small enclosed volumes $V$. The global minimum for the axisymmetric
case enclosing small volumes corresponds to toroidal surfaces. This situation considered here resemble in many ways the classical Goldschmidt discontinuous minimal
area surface of revolution  limited by two coaxial rings separated by a distance $\ell$ \cite{Soap,Gold}. In our case, for $V$ below a critical value, we have two toroidal minimal area surfaces 
enclosing a given volume. One of them has area smaller than the corresponding lens-shaped surface, while the other has
a greater superficial area. In the Goldschmidt case, for $\ell$ below a certain critical value, we have always two 
minimal surfaces (catenoids), but only one of them can effectively have a total superficial area smaller than the discontinuous
Goldschmidt solution. In fact, the diagram Area $\times\  \ell$ for the catenoids is very similar to our 
Area $\times$ Volume curve in Fig. \ref{fig2}.

\begin{figure}[ht]
\includegraphics[scale=0.65]{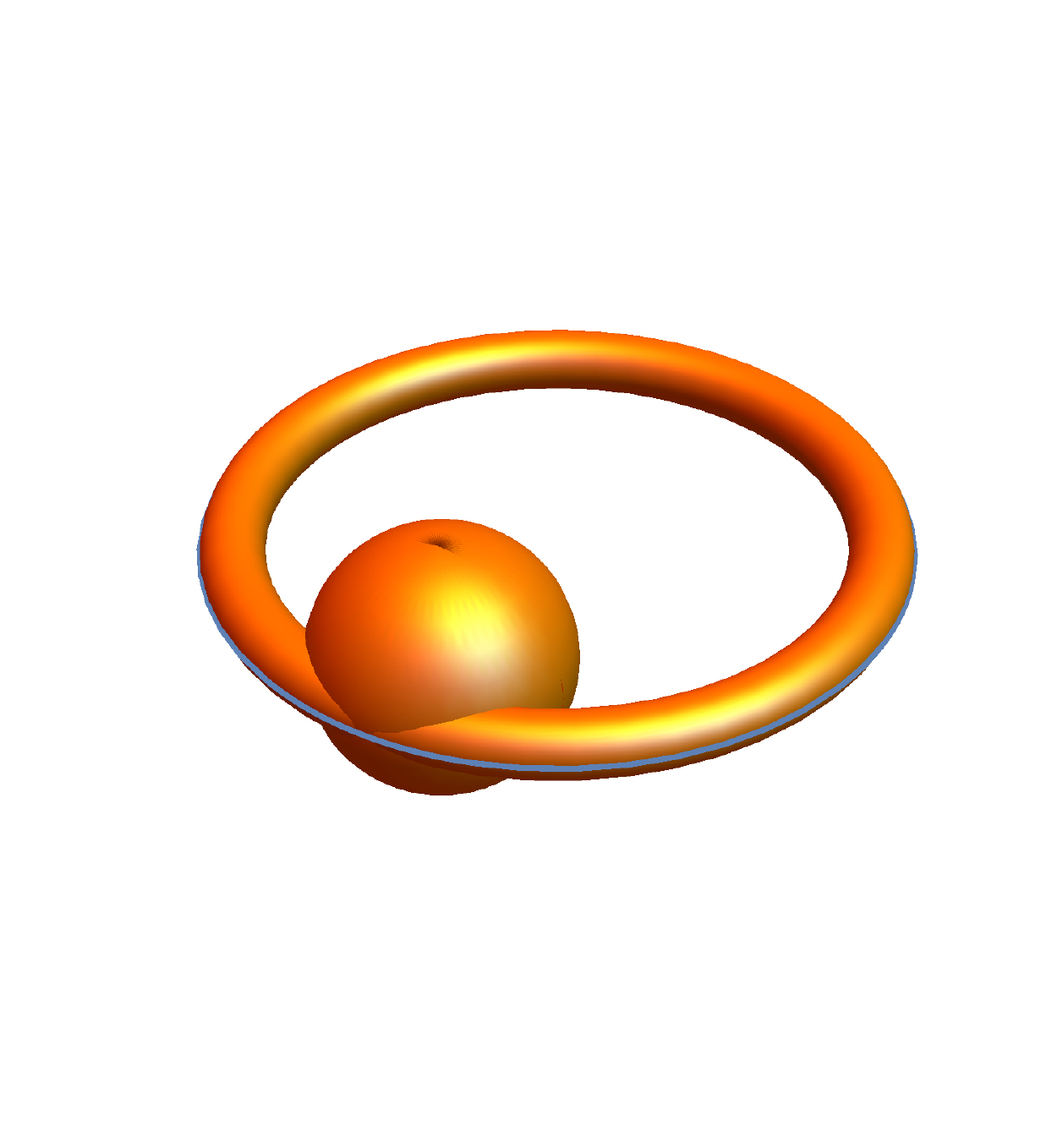}
\vspace{-2cm}
\caption{Sphere with a toroidal handle: by shrinking the handle smaller radius, we have a surface with fixed external perimeter
$L$, enclosing a volume $V<\alpha L^3/6\pi^2$, and with area $A$ smaller than our axisymmetric toroidal surface. Axisymmetric surfaces
are not global solutions for the problem.
}
\label{fig4}
\end{figure}

Finally, we cannot ignore that a doughnut-shaped minimal area surface will be ultimately unstable due to phenomena 
like shrinking \cite{Theory,PFN} and Plateau-Rayleigh \cite{PR,PR1}   instabilities. This means that, for instance, if one deforms a bubble axisymmetricaly in
such a way that its equatorial perimeter $L$ is enlarged while its volume $V$ is kept constant, the bubble will be inexorably 
destroyed when $L^3>6\pi^2 V/\alpha$.  If a doughnut-shaped bubble is formed,
 it will probably break apart in smaller daughter  bubbles
(see, for similar behavior in another context, \cite{Nature}).
It is not difficult to envisage a non-axisymmetric surface with fixed external perimeter $L$, enclosing a volume $V<\alpha L^3/6\pi^2$ and with area $A$ smaller
than the area of our toroidal surface. Consider, for instance, a  sphere with a handle as shown in Fig. \ref{fig4}.
The external handle guarantees the constant perimeter constraint. By shrinking its smaller radius, its contribution for the total $V$ and $A$ will be
arbitrarily small, and hence they will correspond to the sphere, which is known to be the global solution for the problem and will certainly 
encapsulate a given volume $V$ with surface area $A$ smaller than our doughnut-shaped surface. This kind of non-axisymmetric surface might
  arise from a Plateau-Rayleigh instability, where perturbations with wavelength greater than the smaller dimension of the toroidal surface could
  grow exponentially until disrupt it. 
  
 The situation is more
intricate, however, if one keeps the perimeter curve fixed, preventing in this way the formulation  of an axisymmetric problem from the
beginning. By similar arguments,
we also expect genus 1  minimal area surfaces  for small $V/L^3$ in this case, but the value of the threshold $\alpha$ can be
different. 
 Despite of being unstable
as minimal area surfaces, doughnut-shaped
structures are quite common in many dynamical situations, ranging from stains left by   a coffee droplet\cite{coffee} to 
extracellular polymeric bacterial coverages\cite{bac}. In particular, toroidal liquid droplets have been obtained by 
exploring some pyroelectric effects on wetting processes \cite{microlens}. 
  We think our result might be useful to shed more light on some of these problems.

\appendix

\section{Numerics}
It is more convenient for our purposes to introduce some dimensionless quantities. From (\ref{e15}), one can introduce the dimensionless
function $F(x)$ 
\beq
\label{A1}
f(\rho) = \frac{1}{\lambda}F(\lambda \rho),
\eeq
with
\beq
\label{A2}
F(x) = \int_{x_{\rm min}}^{x} \frac{d-s^2}{\sqrt{(s^2-x^2_{\rm min})(x^2_{\rm max}-s^2)}}\, ds.
\eeq
It is clear from (\ref{A1}) that the constant $\lambda$ can be absorbed by a global rescaling of the problem. 
The integral (\ref{A2}) can be expressed by means of elliptical integrals, see, for instance, formulas 9 and 7 in
the sections 3.152 and 3.153, respectively, of \cite{Grad}. However, the expressions result rather
cumbersome for our manipulations, and we chose to solve (\ref{A2}) numerically. 
The integrand diverges for $x=x_{\rm min}$ and $x=x_{\rm max}$,
but the divergence is integrable and it can be easily circumvented, for instance, by introducing the new variable
$u^2=s^2-x_{\rm min}^2$. The singularity for $x=x_{\rm max}$ can be eliminated analogously.

The equatorial 
perimeter will correspond to the external radius of the doughnut-shaped solution, {\em i.e.}, to the point
$\rho_*=\lambda x_*$ such that 
$F(x_*)=0$. The value of $x_*$ can be determined accurately from (\ref{A1}) by using a Newton-Rapson scheme. 
The equatorial perimeter will be given by
\beq
\label{A3}
L = 2\pi\frac{x_*}{\lambda}.
\eeq
The volume enclosed by the doughnut-shaped surface will be given by
\beq
\label{A4}
V = \frac{4\pi}{\lambda^3} \int_{x_{\rm min}}^{x_*} s F(s)\, ds,
\eeq
and its surface area reads
\beq
\label{A5}
A = \frac{8\pi}{\lambda^2} \int_{x_{\rm min}}^{x_*}  \frac{s^2}{\sqrt{(s^2-x^2_{\rm min})(x^2_{\rm max}-s^2)}}\, ds.
\eeq
The  Area $\times$ Volume diagram of (\ref{fig2}) is constructed from (\ref{A4}) and (\ref{A5})  by varying $d$ while keeping
 $L$ given by (\ref{A3}) fixed.

\section*{Acknowledgements}
AS thanks FAPESP (grant 2013/09357-9) and CNPq (grants 304378/2014-3 and 441349/2014-5) 
for the financial support, Ricardo Mosna for enlightening
discussions, and the anonymous referees for useful suggestions.

\end{document}